\documentclass[12pt,english]{article}
\usepackage[T1]{fontenc}
\usepackage[latin9]{inputenc}
\usepackage{geometry}
\usepackage{setspace}
\makeatletter

\newcommand{\noun}[1]{\textsc{#1}}

\newcommand{\lyxaddress}[1]{
\par {\raggedright #1
\vspace{1.4em}
\noindent\par}
}

\makeatother

\usepackage{babel}
\begin{document}

\title{MOND is unnecessary}

\maketitle

\lyxaddress{T.R. Mongan}

\lyxaddress{84 Marin Avenue, Sausalito, California 94965, USA; tmongan@gmail.com}
\begin{abstract}
Dark matter seems to account for flat velocity curves in spiral galaxies,
with further evidence for dark matter from observations of the colliding
\textquotedblleft{}bullet cluster\textquotedblright{} galaxies 1E0657-56.
However, the baryonic Tully-Fisher relation and the mass discrepancy-acceleration
(or V$_{observed}$/V$_{Newtonian}$) relation have been cited (arXiv:1112.3960)
as ``challenges for the $\Lambda CDM$ model.'' MOND (MOdified Newtonian
Dynamics), a modified law of gravity involving an acceleration threshold
$a_{0}\approx1.2\times10^{-8}$cm/sec$^{2}$, is invoked in arXiv:1112.3960
to account for\emph{ }those relations.

This note shows that the HLSS model in arXiv:1301.0304, employing
the holographic principle within the standard $\Lambda$CDM paradigm,
readily accounts for both the MOND acceleration and\emph{ }the (V$_{observed}$/V$_{Newtonian}$)
relation. And, after first posting this note, I learned that Man Ho
Chan (arXiv:1310.6801) previously reached the same conclusion using
a dark matter based analysis independent of the holographic approach
used in this paper. These results indicate that the MOND hypothesis
is unnecessary.
\end{abstract}
Dark matter seems to account for flat velocity curves in spiral galaxies,
and observations of the colliding \textquotedblleft{}bullet cluster\textquotedblright{}
galaxies 1E0657-56 provide further evidence for dark matter. However,
the baryonic Tully-Fisher relation and the mass discrepancy-acceleration
(or V$_{observed}$/V$_{Newtonian}$) relation have been cited \cite{key-1}
as ``challenges for the $\Lambda CDM$ model.'' Ref. 1 invokes MOND
(MOdified Newtonian Dynamics), a modified law of gravity involving
an acceleration threshold $a_{0}\approx1.2\times10^{-8}$cm/sec$^{2}$,
to account for\emph{ }those relations. In contrast, the results below
show that the holographic large scale structure (HLSS) model \cite{key-2}
developed within the $\Lambda$CDM paradigm and employing the holographic
principle based on thermodynamics and general relativity \cite{key-3},
accounts for the MOND acceleration threshold, the baryonic Tully-Fisher
relation, and the mass discrepancy-acceleration (V$_{obseerved}$/V$_{Newtonian}$)
relation.

In the holographic large scale structure (HLSS) model \cite{key-2},
galaxies with total mass $M_{g}$ inhabit spherical holographic screens
with radius $R_{S}=\sqrt{\frac{M_{g}}{0.183g/cm^{2}}}$ if the Hubble
constant $H_{0}=$ 67.8 km sec$^{-1}$Mpc$^{-1}.$ The HLSS model
considers galactic matter density distributions $\rho(r)=\frac{M_{g}}{4\pi R_{S}r^{2}}$,
where $r$ is the distance from the galactic center. The spherical
isothermal halo of dark matter, with radius $R_{S}$ and mass \noun{$M_{DM}=0.84M_{g}$,
}has\noun{ }density distribution $\rho_{DM}(r)=\frac{M_{DM}}{4\pi R_{S}r^{2}}$
so the dark matter mass within radius $R$ is $\frac{R}{R_{S}}M_{DM}.$
There is no singularity in the galactic matter density distribution
$\rho(r)=\frac{M_{g}}{4\pi R_{S}r^{2}}$ because mass inside a core
volume of radius $R_{c}$ at the galactic center is concentrated in
a central black hole with mass $M_{CBH}=\frac{R_{c}}{R_{S}}M_{g}$
\cite{key-2}. Radial acceleration at radius $R$ due to dark matter
is then $a_{DM}=\frac{G}{R^{2}}($$\frac{R}{R_{s}}$)$M_{DM}$. At
radii $R$ sufficiently distant from the galactic center that total
baryonic mass of the galaxy $M_{B}=0.16M_{g}$ can be treated as concentrated
at the galactic center, Newtonian radial acceleration resulting from
baryonic matter is $a_{B}=\frac{GM_{B}}{R^{2}}.$ The radius $R_{\gamma}$
where $a_{DM}=a_{B}$ is found from
\[
\frac{G}{R_{\gamma}^{2}}\left(\frac{R_{\gamma}}{R_{S}}\right)M_{DM}=\frac{GM_{B}}{R_{\gamma}^{2}}.
\]
 Since \noun{$M_{DM}=0.84M_{g}$ }and $M_{B}=0.16M_{g}$, $R_{\gamma}=0.19R_{S}$,
and at that radius $a_{DM}=a_{B}=a_{0}=5.4\times10^{-8}$cm/sec$^{2}$,
consistent with the MOND estimate $a_{0}\approx1.2\times10^{-8}$cm/sec$^{2}$.

Another indication that the MOND acceleration $a_{0}\approx1.2\times10^{-8}$cm/sec$^{2}$
is a natural scale in the dark matter based HLSS model involves the
situation at the radius $R_{S}$ of the spherical holographic screen.
Then the Newtonian assumption that total galactic mass can be treated
as concentrated at the galactic center is certainly justified. There,
the sum of radial acceleration from dark matter and from baryonic
matter is
\[
a_{S}=\frac{GM_{DM}}{R_{S}^{2}}+\frac{GM_{B}}{R_{S}^{2}}=\frac{G}{R_{S}^{2}}(M_{DM}+M_{B}).
\]
 But $M_{DM}+M_{B}=M_{g}=0.183R_{S}^{2}$, so $a_{S}=0.183G=1.2\times10^{-8}$cm/sec$^{2}$,
equal to the estimated MOND acceleration $a_{0}\approx1.2\times10^{-8}$cm/sec$^{2}$.

Turning to the (V$_{observed}$/V$_{Newtonian}$) relation, tangential
velocity $V$ at radius $R$ is related to radial acceleration $a_{r}$
by $V^{2}=Ra_{r}$. So, the ratio (V$_{observed}$/V$_{Newtonian}$)
is approximately
\[
\left(\frac{V_{observed}}{V_{Newtonian}}\right)^{2}\approx\frac{R(a_{DM}+a_{B})}{Ra_{b}}
\]
resulting in
\[
\frac{V_{observed}}{V_{Newtonian}}=\sqrt{1+\frac{a_{DM}}{a_{B}}}=\sqrt{1+\frac{RM_{DM}}{R_{S}M_{B}}}.
\]
Then, when $R\ll0.19R_{s}$, $\frac{V_{observed}}{V_{Newtonian}}\approx1$.
Next, using
\[
\frac{a_{0}}{a_{B}}=\frac{M_{DM}}{0.19M_{B}}(\frac{R}{R_{S}})^{2}
\]
 and 
\[
\frac{R}{R_{S}}=\sqrt{\frac{0.19M_{B}}{M_{DM}}}\sqrt{\frac{a_{0}}{a_{B}}}
\]
yields 
\[
\frac{V_{observed}}{V_{Newtonian}}=\sqrt{1+\frac{RM_{DM}}{R_{S}M_{B}}}=\sqrt{1+\sqrt{\frac{0.19M_{DM}}{M_{B}}}\sqrt{\frac{a_{o}}{a_{B}}}}.
\]
When $a_{B}\ll a_{0},$ $\frac{a_{o}}{a_{B}}\gg1$ and
\[
\frac{V_{observed}}{V_{Newtonian}}=\sqrt{1+0.998\sqrt{\frac{a_{o}}{a_{B}}}}\approx\left(\frac{a_{0}}{a_{B}}\right)^{\frac{1}{4}}
\]
 Since $\left(\frac{V_{observed}}{V_{Newtonian}}\right)^{4}=\frac{a_{0}}{a_{B}}$
when $a_{B}\ll a_{0}$, using $a_{B}=\frac{GM_{B}}{R^{2}}$ and $V_{Newtonian}=\sqrt{\frac{GM_{B}}{R^{2}}}$
gives 
\[
V_{observed}^{4}=\left(\frac{GM_{B}}{R^{2}}\right)^{2}\left(\frac{R^{2}}{GM_{B}}\right)a_{0}=GM_{B}a_{0},
\]
 also known as the baryonic Tully-Fisher relation. 

Finally, with Hubble constant $H_{0}=$ 67.8 km sec$^{-1}$Mpc$^{-1}$,
the cosmological constant $\Lambda=1.12\times10^{-56}$cm$^{-2}$,
and the accelerations $cH_{0}=6.6\times10^{-8}$cm/sec$^{2}$ and
$c^{2}\sqrt{\frac{\Lambda}{3}}=5.5\times10^{-8}$cm/sec$^{2}$ are
both consistent with the acceleration $a_{0}=5.4\times10^{-8}$cm/sec$^{2}$
estimated above.

In conclusion, MOND is not needed to account for the baryonic Tully-Fisher
relation and the mass discrepancy-acceleration (or V$_{observed}$/V$_{Newtonian}$)
relation discussed in Ref. 1. After first posting this note, I learned
that Man Ho Chan previously reached the same conclusion \cite{key-4}
using a dark matter based analysis independent of the holographic
approach used in this paper.


\begin{thebibliography}{1}
\bibitem{key-1}Famaey, B. and Mc Gaugh, S., ``Modified Newtonian
Dynamics (MOND): Observational Phenomenology and Relativistic Extensions,''
Living Reviews in Relativity 15, 10, 2012 (arXiv:1112:3960)

\bibitem{key-2}Mongan, T.R., ``Holography, large scale structure,
supermassive black holes and minimum stellar mass,'' arXiv:1301.0304,
JMP 2, 1544, 2011 and JMP 4, 50, 2013

\bibitem{key-3}Bousso, R., ``The holographic principle,'' Rev.
Mod. Phys. 74, 825, 2002

\bibitem{key-4}Chan, M. H., ``Reconciliation of MOND and Dark Matter
theory,'' Physical Review D88,103501, 2013 (arXiv:1310.6801) \end{thebibliography}
\end{document}